# High-efficiency generation of vectorial holograms with metasurfaces

Tong Liu[1†], Changhong Dai[2†], Dongyi Wang[3*], Lei Zhou[2,4*]

1. Department of Physics, The Hong Kong University of Science and Technology, Clear Water Bay, Kowloon, Hong Kong, China

2. State Key Laboratory of Surface Physics, Key Laboratory of Micro and Nano Photonic Structures (Ministry of Education), Shanghai Key Laboratory of Metasurfaces for Light Manipulation and Department of Physics, Fudan University, Shanghai 200438, China

3. Department of Physics, Hong Kong Baptist University, Kowloon Tong, Hong Kong, China

4. Collaborative Innovation Centre of Advanced Microstructures, Nanjing 210093, China

*Corresponding Authors: Dongyi Wang, Email: phwang@hkbu.edu.hk

Lei Zhou, E-mail: phzhou@fudan.edu.cn

†These authors contributed equally to this work

**ABSTRACT:**

Holography plays a crucial role in optics applications, but it traditionally requires complex setup and bulky devices, being unfavourable for optics integration. While metasurface-based holograms are ultra-compact and easy to realize, holographic images generated are mostly restricted to scalar ones, with a few recent attempts on vectorial holograms suffering from complex meta-structures and low efficiencies. Here, we propose and experimentally demonstrate an efficient meta-platform to generate vectorial holograms with arbitrarily designed wave-fronts and polarization distributions based on ultra-compact metaatoms. Combining Gerchberg–Saxton (GS) algorithm and the wave-decomposition technique, we establish a generic strategy to retrieve the optical property (i.e., the distributions of reflection phase and polarization-conversion capability) of the metasurface to generate a target vectorial holographic image. We next design a series of high-efficiency and deep-subwavelength metaatoms exhibiting arbitrarily designed reflection phases and polarization-conversion capabilities, and experimentally characterize their optical properties. Based on these metaatoms, we finally realize a series of meta-holograms that can generate pre-designed vectorial holographic images upon external illuminations, and experimentally characterize their working performances. Our work provides a high-efficiency and ultra-thin platform to generate vectorial holographic images, which can find many applications in on-chip photonics.

# Introduction

Holography, an optical technology to record and reconstruct a target light field, has attracted intensive attention recently for its many applications in practice such as imaging, virtual reality, data encrypting and storage [1–3]. The traditional approach to generate a hologram requires complex optics interference setup and the generated device usually exhibits a bulky size [4–6], all being highly unfavourable in integration optics. Such limitation is only partially addressed after the Gerchberg–Saxton (GS) algorithm was used to retrieve the optical property of a hologram in computer without doing realistic interference experiments. Moreover, holographic images generated by these approaches are typically scalar ones, which exhibit homogeneous distributions of polarizations.

Metasurfaces, ultra-thin metamaterials composed by subwavelength planar microstructures (e.g. meta-atoms) arranged in certain sequences, have emerged as a powerful platform to manipulate light waves. Many fascinating physical effects were demonstrated based on metasurfaces, including polarization control [7–11], surface-wave coupling [12–14], meta-lensing [15–18], cloaking [19] and so on [20–23]. In particular, choosing meta-atoms exhibiting reflection/transmission phases in accordance with those retrieved from the Gerchberg–Saxton (GS) algorithm, one can construct a metasurface which, under external illumination, capable of generating the pre-designed holographic image [24–28]. Compared to conventional holography technology, such meta-holograms are flat, ultra-compact, easy to realize, and exhibit better lateral resolutions, all being high desired for optics integration. Despite of exciting progresses achieved on meta-holograms, however, most of holographic images realized are scalar ones exhibiting homogeneous polarization distributions. Although a limited attempts were made to generate *vectorial* holographic images exhibiting pre-designed inhomogeneous polarization distributions [29–32], the recording meta-holograms suffer from the issues of low efficiencies and limited lateral resolutions, as the adopted unit phase-bits are constructed with multiple resonators.

In this work, we propose a generic approach to realize *vectorial* holographic images based on high-efficiency metasurfaces exhibiting high lateral resolutions. We first establish efficient logics to retrieve the wave-scattering property of a metasurface for generating a target vectorial image, combining GS algorithm and wave-decomposition technique. In contrast to conventional meta-holograms, here the vectorial meta-holograms exhibit not only inhomogeneous distributions of scattering phases and but also certain distributions of polarization-conversion capabilities. We then design a series of subwavelength single-structure meta-atoms exhibiting freely tailored scattering Jones-matrices, and experimentally characterize their wave-scattering properties. With these building blocks at hand, we finally design and fabricate several meta-holograms and experimentally demonstrate that they can generate pre-designed complex *vectorial* holographic images, upon external illuminations at the wavelength of 1064 nm.

# Results

## Generic strategy for designing vectorial meta-holograms

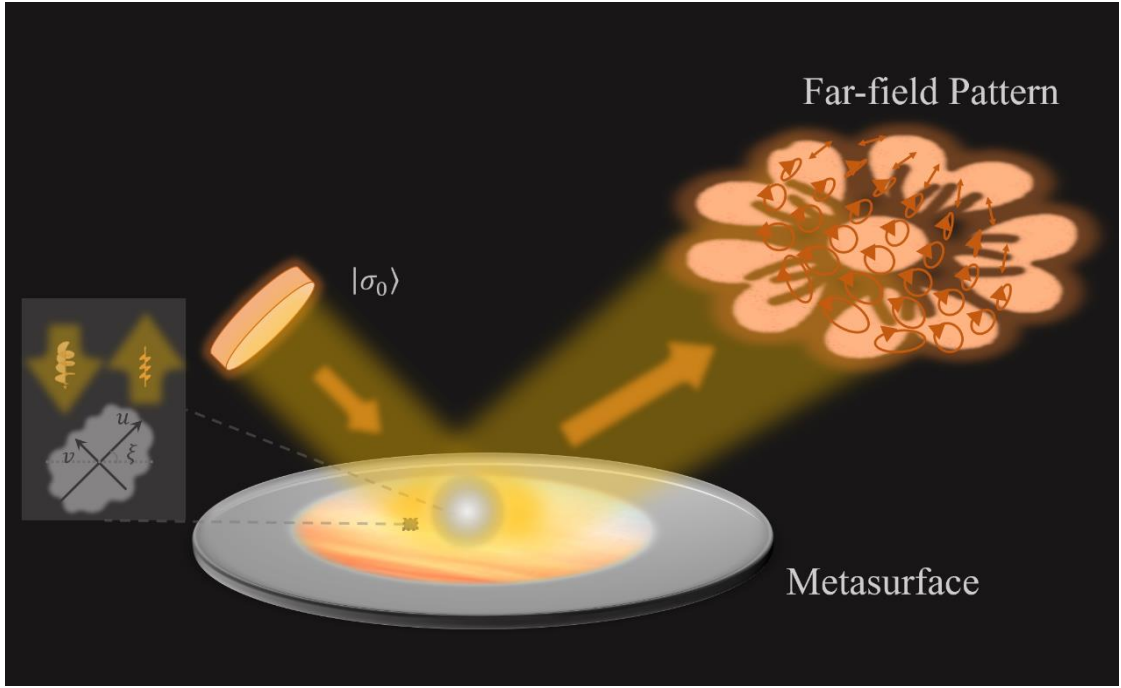

**Fig. 1 Schematics of vectorial hologram generation with simultaneous manipulation of far-field IDs and LPDs based on metasurfaces.** Employing anisotropic meta-atoms carefully arranged on the meta-surface designed by GS algorithm, target IDs and LPDs on the hologram imaging plane without any specific symmetry can be realized simultaneously. The left inset illustrates a meta-atom lying along local axes $(\hat{u}, \hat{v})$, which could convert the incident light into a desired local polarization with arbitrary additional phase.

We now establish a generic strategy to retrieve the optical property of a metasurface, which, as shined by a homogeneously polarized light beam with a polarization $|\sigma_0\rangle$, can generate a target vectorial holographic image in the far field exhibiting tailored intensity and polarization distributions on its wave front (see Fig. 1). To define a far-field vectorial image, it is customized to decompose the beam into **k**-space and use $A_{\text{tar}}^{\text{h}}(\vec{k}), |\sigma_{\text{tar}}^{\text{h}}(\vec{k})\rangle$ to describe its intensity and polarization distributions in the **k**-space. We note that $A_{\text{tar}}^{\text{h}}(\vec{k})$ and $|\sigma_{\text{tar}}^{\text{h}}(\vec{k})\rangle$ precisely represent the intensity and polarization distributions of the field pattern on the focal plane of a lens employed to image the vectorial beam. Therefore, our task is to find an algorithm to retrieve the near-field (NF) scattering property of a metasurface, represented by its distributions of phases/polarizations of waves scattered at different position **r**, from the far-field vectorial image described by $A_{\text{tar}}^{\text{h}}(\vec{k})$ and $|\sigma_{\text{tar}}^{\text{h}}(\vec{k})\rangle$.

To proceed, we rewrite the polarization distribution of the target vectorial beam as $|\sigma_{\text{tar}}^{\text{h}}(\vec{k})\rangle = \begin{pmatrix} e^{-i\Psi_{\text{tar}}^{\text{h}}(\vec{k})/2} \cos(\Theta_{\text{tar}}^{\text{h}}(\vec{k})/2) \\ e^{+i\Psi_{\text{tar}}^{\text{h}}(\vec{k})/2} \sin(\Theta_{\text{tar}}^{\text{h}}(\vec{k})/2) \end{pmatrix}$, where $\{\Theta_{\text{tar}}^{\text{h}}(\vec{k}), \Psi_{\text{tar}}^{\text{h}}(\vec{k})\}$ is a specific point on Poincares' sphere representing the polarization state of the wave with a wave-vector $\vec{k}$.

It is worth noting that all kinds of polarizations can be projected to circular bases composed by left circular polarization (LCP) and right circular polarization (RCP), which are commonly utilized as two orthogonal channels. As Fig. 2 (b) illustrate, with simple derivation, the complex vectorial target could be decoupled to these two channels as:

$$\begin{cases} A_{\text{tar},+}^{\text{h}}(\vec{k}) = A_{\text{tar}}^{\text{h}}(\vec{k}) \left|\cos \frac{\Theta_{\text{tar}}^{\text{h}}(\vec{k})}{2}\right| \\ A_{\text{tar},-}^{\text{h}}(\vec{k}) = A_{\text{tar}}^{\text{h}}(\vec{k}) \left|\sin \frac{\Theta_{\text{tar}}^{\text{h}}(\vec{k})}{2}\right| \\ \Delta\Phi_{\text{tar}}^{\text{h}}(\vec{k}) = \Psi_{\text{tar}}^{\text{h}}(\vec{k}) \end{cases} \quad (1)$$

Where $A_{\text{tar},\pm}^{\text{h}}(\vec{k})$ denote the target IDs of LCP component and its RCP counterpart on the holographic imaging plane in k-space; and $\Delta\Phi_{\text{tar}}^{\text{h}}(\vec{k})$ denotes the corresponding phase difference between these two components. These three targets are then employed in the Gerchberg–Saxton algorithm together with a random initial phase. As shown in Fig. 2 (c), after N times of fast Fourier / inverse fast Fourier transforms, we can obtain stable solutions for phase distributions of LCP and RCP components on the near-field metasurface plane, denoting $\Phi_+^{\text{m}}(\vec{r})$ and $\Phi_-^{\text{m}}(\vec{r})$ respectively. Corresponding nearfield phase targets for generation of the vectorial holography are thus defined by:

$$\begin{cases} \Phi_{\text{tar}}^{\text{m}}(\vec{r}) = \frac{\Phi_+^{\text{m}}(\vec{r}) + \Phi_-^{\text{m}}(\vec{r})}{2} \\ \Theta_{\text{tar}}^{\text{m}}(\vec{r}) = 2 \arctan \frac{\sqrt{I_-}}{\sqrt{I_+}} \\ \Psi_{\text{tar}}^{\text{m}}(\vec{r}) = \Phi_-^{\text{m}}(\vec{r}) - \Phi_+^{\text{m}}(\vec{r}) \end{cases} \quad (2)$$

where $I_\pm = \iint \left(A_{\text{tar},\pm}^{\text{h}}(\vec{k})\right)^2 d^2\vec{k}$ is the intensity of two channels on the whole hologram imaging plane, $\Phi_{\text{tar}}^m(\vec{r})$ is the phase distribution and $\{\Theta_{tar}^m(\vec{r}), \Psi_{tar}^m(\vec{r})\}$ denote LPDs by $|\sigma_{\text{tar}}^m(\vec{r})\rangle = \begin{pmatrix} e^{-i\Psi_{\text{tar}}^m(\vec{r})/2} \cos(\Theta_{\text{tar}}^m(\vec{r})/2) \\ e^{+i\Psi_{\text{tar}}^m(\vec{r})/2} \sin(\Theta_{\text{tar}}^m(\vec{r})/2) \end{pmatrix}$ on the meta-surface plane.

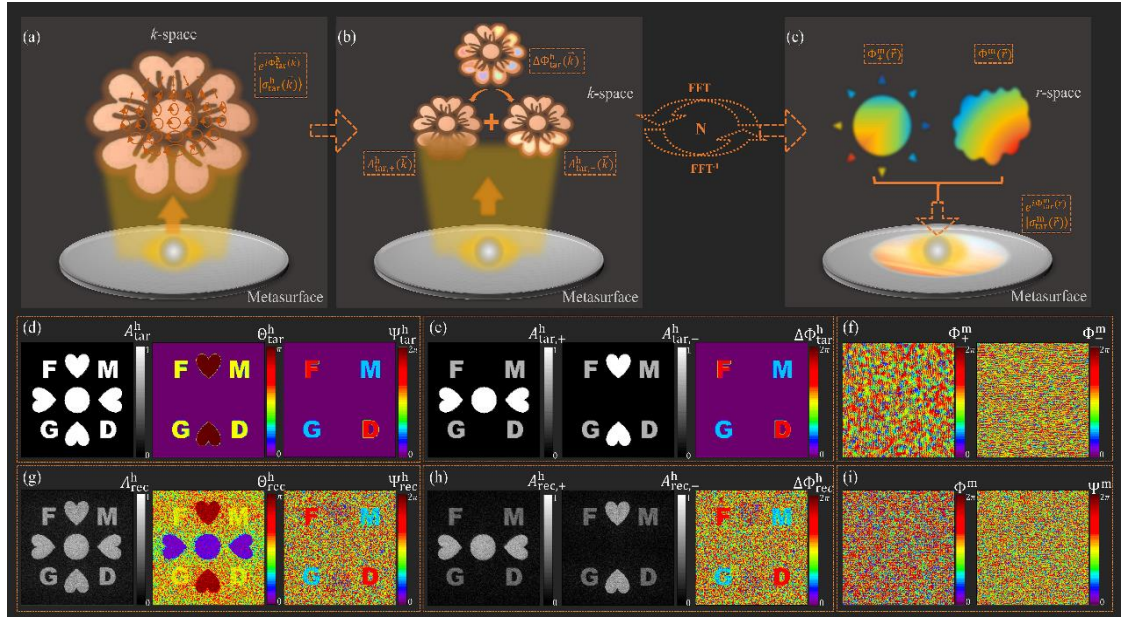

**Fig. 2 Design process for generation of arbitrary vectorial holography based on the generalized GS algorithm and an example designed following this strategy, together with correspondingly reconstructed results.** Flow chart of GS algorithm for vectorial holography generators' design, first step (a) setting target IDs and LPDs on the hologram imaging plane, denoting $A_{\text{tar}}^{\text{h}}(\vec{k})$ and $|\sigma_{\text{tar}}^{\text{h}}(\vec{k})\rangle$; second step (b) decoupling target IDs and LPDs to amplitudes of two orthogonal components $A_{\text{tar},\pm}^{\text{h}}(\vec{k})$ and their phase difference $\Delta\Phi_{\text{tar}}^{\text{h}}(\vec{k})$; third step (c) employing generalized GS algorithm to calculate amplitudes of two orthogonal components $\Phi_{\pm}^{m}(\vec{r})$, then obtaining phase and LPDs on the metasurface plane $\Phi_{\text{tar}}^{m}(\vec{r})$ and $|\sigma_{\text{tar}}^{m}(\vec{r})\rangle$ correspondingly. An example illustrating (d) IDs and LPDs targets denoting $\{A_{\text{tar}}^{\text{h}}, \Theta_{\text{tar}}^{\text{h}}, \Psi_{\text{tar}}^{\text{h}}\}$ on the imaging plane; (e) calculated targets of two orthogonal components denoting $\{A_{\text{tar},\pm}^{\text{h}}, \Delta\Phi_{\text{tar}}^{\text{h}}\}$ on the imaging plane; (f) amplitudes of two orthogonal components denoting $\Phi_{\pm}^{m}(\vec{r})$ calculated by generalized GS algorithm; (i) correspondingly obtained phase and LPDs on metasurface denoted by $\{\Phi_{\text{tar}}^{m}, \Psi_{\text{tar}}^{m}\}$ and $\Theta_{\text{tar}}^{m} = 0.44\pi$ (const.); (h) reconstructed far-field amplitudes and phase difference of two orthogonal components denoted by $\{A_{\text{rec},\pm}^{\text{h}}, \Delta\Phi_{\text{rec}}^{\text{h}}\}$, and (g)

reconstructed far-field IDs and LPDs denoted by $\{A_{rec}^h, \Theta_{rec}^h, \Psi_{rec}^h\}$.

To better illustrate the farfield-to-nearfield design logics, we numerically calculated the generation of a complex vectorial holography as an example. As shown in Fig.2 (d), four characters "F", "D", "M", "G" together with four heart shapes and a circle are set as ID targets on the imaging plane. As for LPDs, all the characters are set to carry LPs and the rest are set to carry CPs. The LPs for "F", "D" are oriented at an 135° polarization angle, while their "M", "G" counterparts are oriented perpendicularly at 45°; the lateral heart-shapes and the circle carry even LCP and the longitudinal ones carry even RCP. These targets are denoted by $\{A_{tar}^h(\vec{k}), \Theta_{tar}^h(\vec{k}), \Psi_{tar}^h(\vec{k})\}$ on the holographic imaging plane in *k*-space. By decomposing the targets to LCP and RCP channels, the amplitudes of these two orthogonal components and their phase difference $\{A_{tar,\pm}^h(\vec{k}), \Delta\Phi_{tar}^h(\vec{k})\}$ are calculated and demonstrated accordingly in Fig.2 (e), which are then repeatedly substituted into the generalized GS algorithm. After N times' iteration, the stable solution for phase distributions of LCP and RCP components $\Phi_\pm^m(\vec{r})$ on the metasurface plane are obtained [Fig. 2 (f)], thus $\Theta_{tar}^m = 0.44\pi$ (const.) and $\{\Phi_{tar}^m(\vec{r}), \Psi_{tar}^m(\vec{r})\}$ can be derived from Eq. 2 [Fig. 2 (i)]. With $\{\Phi_{tar}^m(\vec{r}), \Theta_{tar}^m(\vec{r}), \Psi_{tar}^m(\vec{r})\}$ at hand, amplitudes of LCP/RCP components and their phase difference $\{A_{rec,\pm}^h(\vec{k}), \Delta\Phi_{rec}^h(\vec{k})\}$ are reconstructed in the far-field *k*-space[Fig. 2 (h)]. The IDs and LPDs on holographic imaging plane $\{A_{rec}^h(\vec{k}), \Theta_{rec}^h(\vec{k}), \Psi_{rec}^h(\vec{k})\}$ are correspondingly reconstructed as shown in Fig. 2(g), which is directly comparable to the initial targets $\{A_{tar}^h(\vec{k}), \Theta_{tar}^h(\vec{k}), \Psi_{tar}^h(\vec{k})\}$, being a solid proof for the feasibility of our design strategy. Following this inverse-design process employing farfield-to-nearfield logics and full-matrix phases, arbitrary vectorial holography can be realized as long as appropriate meta-atoms exhibiting desired $\{\Phi_{tar}^m(\vec{r}), \Theta_{tar}^m(\vec{r}), \Psi_{tar}^m(\vec{r})\}$ can be precisely designed.

**Basic meta-atoms and experimental characterizations**

So how to realize such $\{\Phi_{tar}^m(\vec{r}), \Theta_{tar}^m(\vec{r}), \Psi_{tar}^m(\vec{r})\}$ with carefully designed metasurface? A set of anisotropic meta-atoms is demanded, each could convert the incident polarization $|\sigma_0\rangle$ into arbitrary $|\sigma_{tar}^m(\vec{r})\rangle$ and provide an extra phase $\Phi_{tar}^m(\vec{r})$ simultaneously, as shown in the inset of Figure 1 (a). In the following sections, a Metal-Insulator-Metal (MIM) tri-layer meta-atom is employed, as shown in the inset of Figure 3 (b). When illuminated by an incident light, anti-parallel currents are induced on the top metallic structure and bottom substrate, thus form a magnetic resonance mode. With a rotatable cross metallic structure on the top, two perpendicular resonant mode are simultaneously excited, and thus provide enough manipulating DOFs. In this work, the metallic structure and substrate are set to be Ag, while the insulator is $SiO_2$, thickness being $h_{Str} = 30$nm; $h_{Ins} = 125$nm; $h_{Sub} = 125$nm respectively. The top bar width is $w = 80$nm, periodicities along $x, y$ direction are $P_x = P_y = 600$nm. Bar lengths along two local axes $L_u$, $L_v$ are variable, thus corresponding phases $\Phi_u$, $\Phi_v$ are tuneable; together with the rotation angle $\xi$, the phase and LPDs on metasurface $\Phi_{tar}^m(\vec{r})$ and $|\sigma_{tar}^m(\vec{r})\rangle$ can be freely modulated by:

$$\begin{cases} \Phi_{tar}^m(\vec{r}) = \Phi_{Res}(\vec{r}) + \sigma_0 \Phi_{Geo}(\vec{r}) \\ \Theta_{tar}^m(\vec{r}) = \sigma_0 \left(\Delta\Phi - \frac{\pi}{2}\right) + \frac{\pi}{2} \\ \Psi_{tar}^m(\vec{r}) = 2\xi(\vec{r}) - \sigma_0 \frac{\pi}{2} \end{cases} \quad (3)$$

when $\Delta\Phi \neq \pm\pi$; and $\Theta_{tar}^m(\vec{r}) = \pi\sigma_0/2 + \pi/2$, $\Psi_{tar}^m(\vec{r}) = 0$ when $\Delta\Phi = \pm\pi$. Here, $\Delta\Phi = \Phi_v - \Phi_u$ is the phase difference; $\Phi_{Res} = (\Phi_u + \Phi_v)/2 - \pi/4$ when $\Delta\Phi \neq \pm\pi$; $\Phi_{Res} = \arg(e^{i\Phi_u} - e^{i\Phi_v})$ when $\Delta\Phi = \pm\pi$ is the resonant phase modulated by bar-lengths-tuned resonance; $\Phi_{Geo} = \xi$ when $\Delta\Phi \neq \pm\pi$; $\Phi_{Geo} = 2\xi$ when $\Delta\Phi = \pm\pi$ is the geometrical phase merely modulated by rotation angle $\xi$. The dependence of $\Delta\Phi$ and $\Phi_{Res}$ against $L_u$, $L_v$ are calculated numerically by FDTD simulations, as shown in Fig. 3 (a,b). To verify the feasibility of the meta-atom data base, three different meta-atoms labelled No. 1,2,3 are chosen from the phase diagram [Fig. 2 (b)], whose $\Delta\Phi = -3/2\pi, -\pi, 2\pi/3$, thus are supposed to function as quarter waveplate (QWP), half waveplate (HWP) and 2/3 waveplate respectively. Periodically repeating these meta-atoms to form different meta-waveplates and experimentally characterize them, at 45° oriented LP incidence [Fig.2 (c)], the polarizations of reflected light are converted to CP, 135° oriented LP, and EP ones as Fig.2 (d~f) illustrate, well-consistent with theoretical predictions.

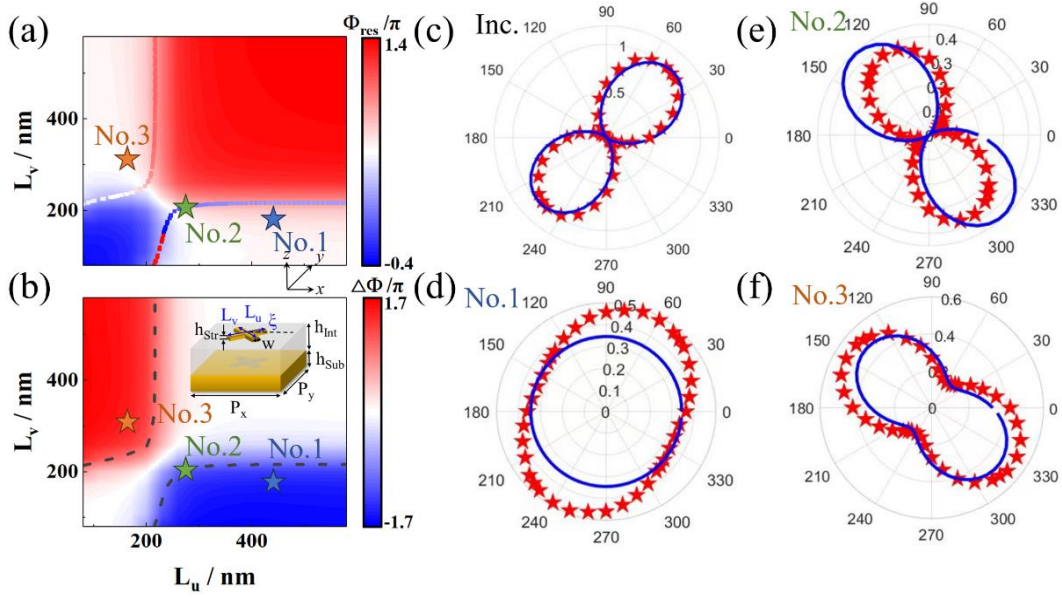

**Fig. 3 Phase diagrams and experimental characterizations of some meta-atoms employed.** Phase diagrams of (a) resonance phase $\Phi_{Res}$ and (b) phase difference $\Delta\Phi$ versus bar-lengths along two local axes $L_u$ and $L_v$ calculated by FDTD simulations at a wavelength of 1064 nm, inset demonstrating schematics of a tri-layer meta-atom employed for hologram designs in the following sections. Fixed geometric parameters marked in the figure are: thickness of metal (Ag) structure $h_{Str} = 30$nm, insulator (SiO$_2$) layer $h_{Ins} = 125$nm and metal (Ag) substrate $h_{Sub} = 125$nm, bar width $w = 80$nm, and periodicities along x, y direction $P_x = P_y = 600$nm. Dashed line denotes the $\Delta\Phi = \pi/2$ contour, No. 1, 2, 3 and stars in blue, green, orange colours denote meta-atoms functioning as quarter, half and 2/3 waveplates which are experimentally characterized. Polarized-filtered intensity patterns of (c) incident light which is a 45° oriented LP one; and light reflected by (d) the meta-quarter waveplate, which is converted to a CP one; (e) the meta-half waveplate, which is converted to a crossed LP one; (f) the meta-2/3 waveplate, which is converted to an EP one.

One thing to notice is that we deliberately design the bar width and dielectric thickness to suppress metallic absorption, thus $|r_{uu}|$ and $|r_{vv}|$ are generally close to 1 [see supplementary information], making approximation of ignoring amplitudes variation in the following sections reasonable. Once $\Phi_{tar}^m(\vec{r})$ and $|\sigma_{tar}^m(\vec{r})\rangle$ determined by GS algorithm according to the target holographic pattern, we can use the data base in this phase diagram together with analytical calculations to conduct geometric parameters' design.

With such generic idea and design strategy, we further verify the platform by designing and experimentally realizing several different vectorial holography generators, as illustrated in the following sections.

## Benchmark Vectorial Holography Generators

In this section, we first employ the generalized vectorial GS algorithm to experimentally demonstrate two simple vectorial holography generations at 1064nm as benchmark cases. These two devices can display simple arrow ID patterns denoting radial and non-cylindrical LPDs respectively on the holographic plane. Following the design process introduced in the last subsection, after *N* times of iterative calculation for convergence, two delinked sets of derived phases on meta-surface, $\Phi_{\pm}^{m}(\vec{r})$ are attained from the optimization process, combing Eqs. (2,3), at LCP incidence, the target parameters on the metasurface plane can be determined as:

$$\begin{cases} \Phi_{Res}(\vec{r}) = \Phi_{+}^{m}(\vec{r}) - \frac{\pi}{4} \\ \Delta\Phi(\vec{r}) = 2\arctan\frac{\sqrt{I_{-}}}{\sqrt{I_{+}}} \\ \xi(\vec{r}) = \frac{\Phi_{-}^{m}(\vec{r}) - \Phi_{+}^{m}(\vec{r})}{2} + \frac{\pi}{4} \end{cases} \quad (4)$$

Thus, geometrical parameters can be designed accordingly with the help of $\{\Phi_{Res}(\vec{r}), \Delta\Phi(\vec{r})\}$ database in Fig. 3 (a,b).

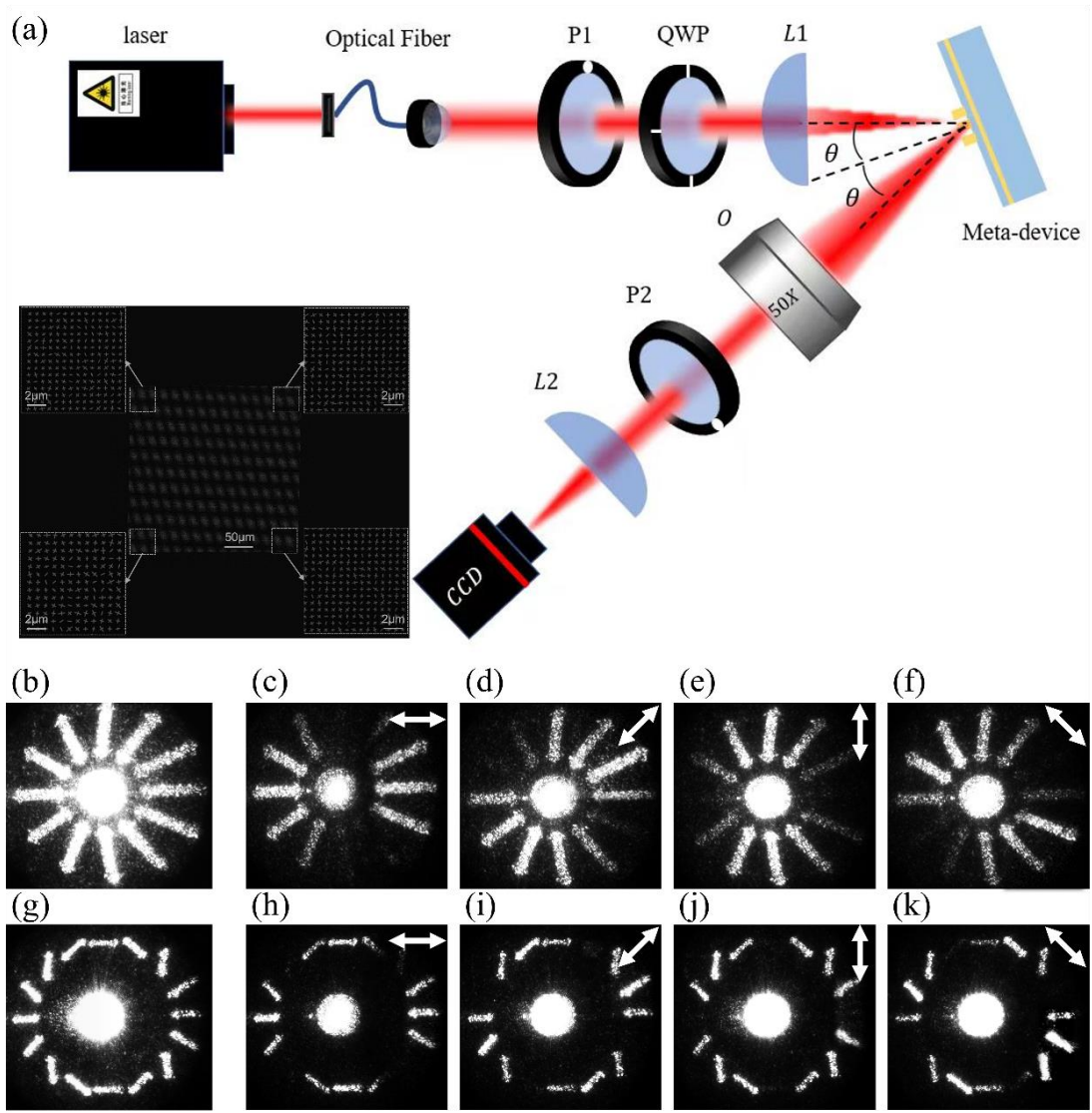

**Fig. 4 Experimental set-up, SEM images and basic demonstrations for Vectorial Holography Generators Based on Metasurfaces.** (a) The experimental set-up for far-field vectorial hologram characterization, P denoting polarizer, QWP denoting quarter wave plate, L denoting lens and O denoting optical lens. The inset illustrates SEM picture and zoomed-in views of the first vectorial holography generator for radially polarized hologram. Experimentally observed (b) radially polarized hologram and polarized-filtered patterns with a rotatable polarizer placed in front of CCD as an analyser, tilted by (c) 0°, (d) 45°, (e) 90°, (f) 135° respectively. Experimentally observed (g) non-cylindrically polarized hologram and polarized-filtered patterns with a rotatable polarizer placed in front of CCD as an analyser, tilted by (h) 0°, (i) 45°, (j) 90°, (k) 135° respectively.

We then fabricated the corresponding sample [the inset of Fig.4 (a)], and experimentally set up a micro-imaging system for characterization of the generated vectorial holography [Fig.4 (a)]. The polarization of the laser beam is converted into an LCP one by the combination of a polarizer P1 and a quarter-wave plate QWP1 tiled at a relative orientation angle of 45°. The meta-device is placed at the focal plane of lens L1 (focal length being 175mm) with an incident angle of $\theta = 22°$. A high numerical aperture (N.A.=0.55) condenser lens O together with another lens L2 (focal length being 100mm) are employed to collect and focus the vectorial holography on to a CCD camera placed at its focal plane. An additional rotatable polarizer P2 is placed in front of the CCD for analysis of LPDs on the holographic plane. As shown in Fig. 4(b), the first vectorial holography generator forms ID with radially-arranged arrows' pattern, whereas the LPDs are supposed to be radius vectorial LP ones just like the pattern indicates. Adding the rotatable analyser, as Fig. 4(c~f) illustrate, the LPDs generated by the first meta-device are LPs parallel to the radial direction, thus intensity zeros occur at different orientations when the polarizer is tiled at different angles, consisting well with the theoretical predictions. As for the second vectorial holography generator, an ID pattern composed by non-cylindrical distributed arrows indicating non-cylindrical LPDs is supposed to be generated, and the corresponding pattern has been explicitly observed in the experiment [Fig. 4(g)]. As seen in Fig. 4(h~k), when rotating the polarization analyser to capture LPD properties experimentally, intensity zeros appear and rotate with the analyser accordingly, living up to the theoretical expectations.

# Complex Vectorial Holography Generators

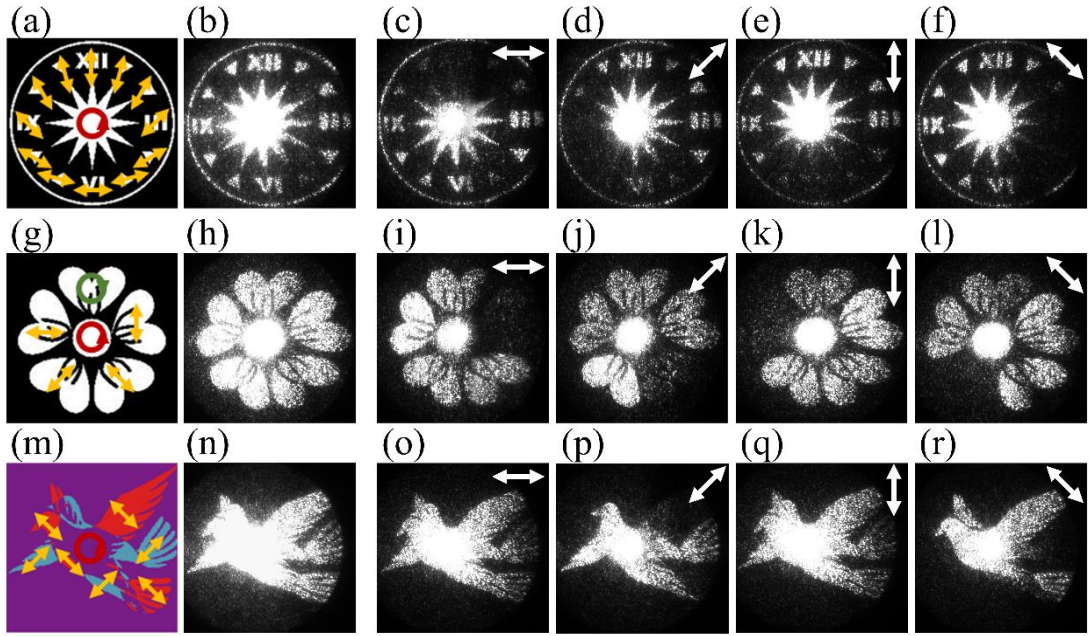

**Fig. 5 Complex Vectorial Holography Generators for completely asymmetric ID and LPD generation Based on Metasurfaces.** Target IDs and LPDs at hologram imaging plane for (a) a vectorial clock; (g) a vectorial flower; (m) a vectorial flying bird. At LCP incidence, experimentally observed (b) vectorial clock hologram and polarized-filtered patterns with a rotatable polarizer placed in front of CCD as an analyser, tilted by (c) 0°, (d) 45°, (e) 90°, (f) 135° respectively. Experimentally observed (h) vectorial flower hologram and polarized-filtered patterns with a rotatable polarizer placed in front of CCD as an analyser, tilted by (i) 0°, (j) 45°, (k) 90°, (l) 135° respectively. Experimentally observed (n) vectorial flying bird hologram and polarized-filtered patterns with a rotatable polarizer placed in front of CCD as an analyser, tilted by (o) 0°, (p) 45°, (q) 90°, (r) 135° respectively. Blue parts in (n) denote 45° oriented LPs, while their red and purple counterparts denote 135° oriented LPs and LCPs respectively.

Encouraged by the success of the benchmark cases, we continue to design three complex vectorial holography generators for formation of holography with IDs and LPDs which are both completely-asymmetric, surpassing the predecessors and demonstrating the full power of our generic design approach. As shown in Fig.5 (a), the third vectorial holography generator is supposed to display a vectorial clock with differently oriented LPs on the azimuthal direction. Thus, a full clock pattern is observed in the experiment when no polarizer is employed as analyser [Fig.5 (b)]; when the analyser is placed in front of the CCD and rotated, we can see different parts on the clock are lighten up in turn (in the opposite direction of the intensity zeros) [Fig.5 (c~f)], as if time is passing by with the tick tock. The fourth vectorial holography generator should be capable to display a five-petal flower [Fig.5 (g)], the centre and five petals of which carry different LPDs, being LCP and RCP for the flower centre and the top pedal, and 0°, 45°, 135°, 90° oriented LPs for the rest pedals along the anti-clockwise direction. As shown in Fig.5 (h~l), the complete flower is observed without the polarization analyser; while different petals are "picked off" when the analyser is employed and tiled at varied angles. For the last vectorial holography generator, we aim at realizing a flying bird when rotate the polarization analyser, thus in Fig. 5(m), the holographic plane pattern is divided into three different categories regarding LPDs, being 45°, 135° oriented LPs and LCPs, denoted by colours of blue, red and purple respectively. With no analyser [Fig. 5(n)] or an analyser tiled at 0°/90° [Fig. 5(o,q)], all patterns are bright; while when the analyser is tiled at 45°, 135°, a bird with a raised head and lowered wings [Fig. 5(p)], and a bird with a lowered head and raised wings [Fig. 5(r)], are observed respectively. The bird "flies" when the analyser is rotated continuously.

The vectorial holograms generated by the meta-platform which breaks the constrains on symmetry and gradual-transition for both IDs and LPDs are unprecedent. All these fancy experimental results consist well with the predictions of our generic vectorial holography design strategy, which firmly proves the powerfulness of the proposed meta-system.

## Conclusions

To conclude, based on brand-new farfield-to-nearfield logics and extended GS algorithm, we have set up a generic platform based on metasurface for high-efficient realization of arbitrary vectorial holography carrying inhomogeneous IDs and LPDs. Under this guideline, employing a generalized vectorial GS algorithm utilizing full-matrix phases, we first experimentally realized two simple vectorial holography generators as benchmark cases, displaying radially distributed arrow ID patterns denoting radius LPDs and non-cylindrical distributed arrow ID patterns denoting non-cylindrical LPDs respectively. Then, three complex vectorial holography generators are designed and experimentally demonstrated, including a vectorial clock ID pattern with asymmetric LPDs which indicates the pass of time when the analyzing polarizer rotates; a vectorial flower ID pattern with piecemeal LPDs (both asymmetric and non-gradual-transited) whose petals are "picked off" respectively when the analyser is oriented at some specific angles; and a vectorial flying bird pattern with completely asymmetric and non-gradual-transited ID pattern and LPDs, displaying a flying bird when the analyser rotates continuously. How to extend this holographic design strategy to even more complex conditions, like adding the angular dispersion/ frequency dispersion/ mode coupling DOFs are valuable future goals.

The proposed versatile, integrated, and highly efficient meta-platform is poised to uncover additional insights into holographic physics in the future. Its capabilities to produce arbitrary asymmetric vectorial holograms with distinct properties could herald a new era of photonic holographic devices. These advancements hold broad application prospective in a wide array of applications, including optical information encryption, anti-counterfeiting measures, data storage, etc.

## Methods

### Numerical simulation

In our finite-difference time-domain simulations, the permittivity of Ag is described by the Drude model $\varepsilon_r(\omega) = \varepsilon_\infty - \frac{\omega_p^2}{\omega(\omega + i\gamma)}$, with $\varepsilon_\infty = 5, \omega_p = 1.367 \times 10^{16} s^{-1}, \gamma = 1.474 \times$

$10^{14} s^{-1}$, obtained by fitting with experimental results. The SiO$_2$ spacer is considered as a lossless dielectric with permittivity $\varepsilon = 2.25$. Additional losses caused by surface roughness and grain boundary effects in thin films as well as dielectric losses are effectively considered in the fitting parameter $\gamma$.

## Sample fabrications

All MIM samples are fabricated using standard thin-film deposition and EBL techniques. In the first step, we sequentially deposit 125 nm – thick Ag and a 125 nm - thick SiO$_2$ dielectric layer onto a silicon substrate using magnetron DC sputtering (Ag) and RF sputtering (SiO$_2$). Then, we lithograph the cross structures with EBL, employing an ~100 nm thick PMMA2 layer at an acceleration voltage of 100 keV. After development in a solution of methyl isobutyl ketone and isopropyl alcohol, a 3 nm Ti adhesion layer and a 30 nm Ag layer are subsequently deposited using thermal evaporation. The Ag patterns are finally formed on top of the SiO$_2$ film after a lift-off process using acetone.

## Experimental setup

We use a near-infrared (NIR) microimaging system to characterize the performance of all designed meta-atoms. A broadband supercontinuum laser (Fianium SC400) source and a fibre-coupled grating spectrometer (Ideaoptics NIR2500) are used in the FF measurements. A beam splitter, a linear polarizer and a CCD are also used to measure the reflectance and analyse the polarization distributions.

# List of abbreviations

degrees of freedom (DOFs)

intensity distribution (ID)

local polarization distribution (LPD)

Gerchberg–Saxton (GS)

circular-polarized (CP)

linear-polarized (LP)

elliptical-polarized (EP)

left circular polarization (LCP)

right circular polarization (RCP)

Scanning electron microscopy (SEM)

Metal-Insulator-Metal (MIM)

near-infrared (NIR)

# Declarations

## Ethics approval and consent to participate

Not applicable.

## Consent for publication

Not applicable.

## Availability of data and materials

The datasets used and/or analyzed during the current study are available from the corresponding author on reasonable request.

## Competing interests

The authors declare that they have no competing interests.

## Funding

This work was funded by National Key Research and Development Program of China (Grant No. 2022YFA1404701), National Natural Science Foundation of China (Grant Nos. 12221004, 62192771) and Natural Science Foundation of Shanghai (Grant Nos. 20JC141460).

# Authors' information

Tong Liu and Changhong Dai contributed equally to this work.

## Authors and Affiliations

Department of Physics, The Hong Kong University of Science and Technology, Clear Water Bay, Kowloon, Hong Kong, China

Tong Liu

State Key Laboratory of Surface Physics, Key Laboratory of Micro and Nano Photonic Structures (Ministry of Education) and Department of Physics, Fudan University, 200433, Shanghai, People's Republic of China

Changhong Dai and Lei Zhou

Department of Physics, Hong Kong Baptist University, Kowloon Tong, Hong Kong, China

Dongyi Wang

Academy for Engineering and Technology, Fudan University, 200433, Shanghai, People's Republic of China

Lei Zhou

Collaborative Innovation Center of Advanced Microstructures, 210093, Nanjing, People's Republic of China

Lei Zhou

# Authors' Contributions

T.L. and C.D. contributed equally to this work. T.L. performed all theoretical derivations and design of holography generators. C.D. fabricated all samples and carried out experimental testing. D.W. and L.Z. conceived the idea and supervised the project. All authors contributed to the discussion and preparation of the manuscript. The authors read and approved the final manuscript.

## Acknowledgements

D. Wang acknowledge support from Prof. Guancong Ma, Department of Physics, Hong Kong Baptist University. L. Zhou acknowledges technical support from the Fudan Nanofabrication Laboratory for sample fabrication.